\documentclass[prl,reprint, twocolumn,showpacs,preprintnumbers,amsmath,amssymb,nofootinbib,floatfix]{revtex4-1} 
\usepackage{graphicx}

\usepackage{mathrsfs}
\usepackage{hyperref}

\usepackage{slashed}

\usepackage{color}

\usepackage{gensymb}

\usepackage{amsmath}
\allowdisplaybreaks[4]

\def \matrix #1 {\left(\begin{array}{cc} #1 \end{array}\right)}

\def\II{\hbox{{1}\kern-.25em\hbox{l}}}

\begin{document}

\title{Renormalization-Group Evolution for the Bottom-Meson Soft Function }

\author{Yong-Kang Huang$^{a}$}
\email{huangyongkang@mail.nankai.edu.cn}

\author{Yao Ji$^{b}$}
\email{corresponding author: yao.ji@tum.de}

\author{Yue-Long Shen$^{c}$}
\email{corresponding author: shenylmeteor@ouc.edu.cn}

\author{Chao Wang$^{d}$}
\email{corresponding author: chaowang@nankai.edu.cn}

\author{Yu-Ming Wang$^{a}$}
\email{corresponding author: wangyuming@nankai.edu.cn}

\author{Xue-Chen Zhao$^{a}$}
\email{corresponding author: zxc@mail.nankai.edu.cn}

\affiliation{${}^a$ School of Physics, Nankai University,
Weijin Road 94, Tianjin 300071, P.R. China  \\
${}^b$ Physik Department T31, James-Franck-Stra{\ss}e1, Technische Universit\"{a}t  M\"{u}nchen,
D–85748  Garching,  Germany \\
${}^c$ College of Information Science and Engineering, Ocean University of China,
Qingdao 266100, Shandong, China \\
${}^d$ Department of Mathematics and Physics,
Huaiyin Institute of Technology,
Meicheng East Road 1,  Huaian, Jiangsu 223200, P.R. China }

\date{\today}

\begin{abstract}
\noindent
We determine for the first time the renormalization-group (RG) evolution equation for the $B$-meson soft function
dictating the non-perturbative strong interaction dynamics of the long-distance penguin contributions
to the exclusive  $b \to q \ell^{+} \ell^{-}$ and $b \to q \gamma$ decays.
The distinctive feature of the ultraviolet renormalization of this fundamental distribution amplitude
consists in the novel pattern of mixing positive into negative support for an arbitrary initial condition.
The exact solution to this integro-differential RG evolution equation of the bottom-meson soft function is then derived with
the Laplace transform technique, allowing for the model-independent extraction of the desired asymptotic behaviour
at large/small partonic momenta.
\\[0.4em]

\end{abstract}

\preprint{TUM-HEP-1489/23}

\maketitle

%
\section{Introduction}
%

The light-cone distribution amplitudes (LCDAs) of the bottom-meson 
are the fundamental non-perturbative ingredients for the model-independent description of
exclusive $B$-meson decays into energetic particles. 
These crucial hadronic quantities are in high demand for exploring factorization properties of
a wide range of the non-hadronic $B$-meson decay form factors \cite{Korchemsky:1999qb,Lunghi:2002ju,Bosch:2003fc,Beneke:2011nf,Galda:2022dhp,Beneke:2020fot,Shen:2020hfq,Wang:2021yrr},
and of the semi-leptonic and non-leptonic heavy-hadron decay amplitudes \cite{Beneke:2000wa,Beneke:2005gs,Hill:2004if,Beneke:1999br,Beneke:2000ry,Lu:2022kos}
at leading power in $\Lambda_{\rm QCD}/m_b$.
They also dictate the infrared dynamics of the appropriate $B$-meson-to-vacuum correlation functions
suitable for constructing the desired light-cone sum rules of numerous  bottom-meson decay matrix elements
\cite{Khodjamirian:2005ea,Khodjamirian:2006st,DeFazio:2005dx,DeFazio:2007hw,Li:2009wq,Wang:2015vgv,Lu:2018cfc,Cui:2022zwm,
Gao:2019lta,Wang:2017jow,Gao:2021sav,Cui:2023jiw,Braun:2012kp,Wang:2016qii,Wang:2018wfj,Beneke:2018wjp,Khodjamirian:2023wol}.
Consequently, it has become the top priority to deepen our understanding towards both the non-perturbative behaviours
\cite{Braun:2003wx,Khodjamirian:2020hob,Rahimi:2020zzo,Wang:2019msf} and the perturbative features \cite{Lange:2003ff,Bell:2013tfa,Braun:2014owa,Braun:2019wyx,Lee:2005gza,Feldmann:2014ika,
Galda:2020epp,Liu:2020ydl,Feldmann:2022uok,Feldmann:2023aml,Kawamura:2008vq,Kawamura:2010tj}
of the  $B$-meson LCDAs for the sake of pinning down the theory uncertainties of the exclusive $B$-meson decay observables,
motivated by the ever-increasing precision of the experimental measurements at LHCb and Belle II.

Advancing the field-theoretical computations of the interesting   $B$-meson decay observables
in the perturbative factorization framework necessitates further  the robust  control of
the subleading-power contributions in the  heavy quark expansion.
Unsurprisingly, the higher-twist bottom-meson LCDAs from the non-leading spin projections,
from the transverse motion of quarks and anti-quarks in the leading-twist components,
and from the non-minimal Fock states with additional partonic fields,
defined by the subleading light-cone matrix elements in heavy quark effective theory (HQET) \cite{Kawamura:2001jm,Braun:2015pha,Braun:2017liq,Braun:2018fiz,Descotes-Genon:2009jif,Knodlseder:2011gc,Braun:2022gzl},
will  appear in the theory description of the power-suppressed corrections with the QCD-based methods \cite{Beneke:2003pa,Khodjamirian:2010vf,Khodjamirian:2012rm,Gubernari:2020eft,Piscopo:2023opf}.
However, the intricate soft and collinear strong interaction fluctuations in the subleading-power contributions
to the exclusive $B$-meson decay amplitudes are not necessarily  captured  by the non-local matrix elements
of the composite operators with quark-gluon fields localized on the same light-cone direction \cite{Kozachuk:2018yxf,Melikhov:2019esw,Melikhov:2022wct,Melikhov:2023pet,Belov:2023xqk,Qin:2022rlk}
(see \cite{Benzke:2010js} for  discussions on the inclusive $\bar B \to X_s \gamma$ decay).
An excellent manifestation of the complex infrared structure for the exclusive heavy-hadron decay amplitude
at next-to-leading power can be understood from the soft-collinear factorization formula of the long-distance penguin contribution
to the double radiative $\bar B_{d, s} \to \gamma \gamma$ decays,
which demands an introduction of  the generalized three-particle distribution amplitude
$\Phi_{\rm G}(\omega_1, \omega_2, \mu)$  defined  by the soft matrix element with non-aligned fields \cite{Qin:2022rlk}.
Importantly, this new type of subleading  $B$-meson soft functions will be also indispensable for
QCD calculations of the charm-loop effects  in the flagship electroweak penguin  decay channels
at the LHCb experiment.
Extending the generalized HQET distribution amplitudes to investigations  of  the subleading-power corrections
in  the  semileptonic $\Lambda_b \to \Lambda \ell^{+} \ell^{-}$  decays at large hadronic  recoil \cite{Wang:2008sm,Aslam:2008hp,Wang:2009hra,Mannel:2011xg,Feldmann:2011xf,Wang:2011uv,Boer:2014kda,Wang:2015ndk,Wang:2015ndk}
can be further anticipated (see  for instance \cite{Feldmann:2023plv})
under the influence of the  improved LHCb measurements  on the  branching fraction and angular observables \cite{LHCb:2015tgy}.

Apparently, the eventual factorization formulae of the long-distance penguin contributions
to the exclusive  $b \to q \ell^{+} \ell^{-}$ and $b \to q \gamma$ decays cannot be established without
controlling the renormalization-scale dependence of such generalized $B$-meson distribution amplitudes.
It is the primary objective  of this Letter to derive the renormalization-group (RG) evolution equation of the  soft function
$\Phi_{\rm G}(\omega_1, \omega_2, \mu)$ 
and to present subsequently its exact analytic solution with the Laplace transform technique.
We will then report on a novel observation of the dynamical properties of this generalized distribution amplitude,
in striking contrast to the  three-particle $B$-meson LCDAs \cite{Braun:2017liq},
which can be attributed to the soft-gluon interaction  between the two  Wilson lines in distinct light-cone directions.
Phenomenological implications of the evolution effect due to the one-loop anomalous dimension
will be further discussed with one sample model for  the generalized bottom-meson distribution amplitude.

%
\section{The RG evolution equation}
%

The generalized bottom-meson distribution amplitude $\Phi_{\rm G}$  entering the factorization formula
for the soft-gluon radiative corrections to $\bar B_q \to \gamma \gamma$
is defined by the non-local  HQET matrix element \cite{Qin:2022rlk}
\begin{widetext}
\begin{eqnarray}
&& \langle 0 | (\bar q_{s} S_{n}) (\tau_1 n) \, (S_{n}^{\dagger} \, S_{\bar n})(0) \,
(S_{\bar n}^{\dagger} \, g_s \, G_{\mu \nu} \, S_{\bar n} )(\tau_2 \bar n) \,\,
\bar n^{\nu} \not \! n \gamma_{\perp}^{\mu} \gamma_5 \,
(S_{\bar n}^{\dagger} h_v) (0) | \bar B_v \rangle
\nonumber \\
&& = 2 \, {\cal F}_B(\mu) \, m_B \, \int_{-\infty}^{+\infty} d \omega_1 \, \int_{-\infty}^{+\infty}  d \omega_2 \,
{\rm exp} \left [- i (\omega_1 \tau_1 + \omega_2 \tau_2) \right ] \,
\Phi_{\rm G}(\omega_1, \omega_2, \mu)\,,
\label{definition of phi_G}
\end{eqnarray}
\end{widetext}
where the soft Wilson lines $S_n$ and $S_{\bar n}$ along the distinct light-cone directions of $n$ and  $\bar n$
are introduced to maintain gauge invariance.
To determine the RG  equation of  $\Phi_{\rm G}$,
we first express the renormalized operator ${\cal O}_{\rm G}^{\rm ren}$
in terms of the corresponding  bare operator
\begin{eqnarray}
{\cal O}_{\rm G}^{\rm ren}(\omega_1, \omega_2, \mu) &=&
\int_{-\infty}^{+\infty} d \omega_1^{\prime} \int_{-\infty}^{+\infty}  d \omega_2^{\prime}
Z_{\rm G}(\omega_1, \omega_2, \omega_1^{\prime}, \omega_2^{\prime}, \mu)
\nonumber \\
&& \times \, {\cal O}_{\rm G}^{\rm bare}(\omega_1^{\prime}, \omega_2^{\prime}) \,,
\end{eqnarray}
where ${\cal O}_{\rm G}$ stands for the two-dimensional Fourier transform
of the non-local operator  on the left-hand side of (\ref{definition of phi_G}).
The convolutions in $\omega_{1, 2}^{\prime}$ arise from the fact that the composite operators
with different momentum variables $\omega_{1, 2}$ can mix into each other under the ultraviolet (UV) renormalization.
The renormalization constant $Z_{G}$ calculable in perturbation theory enables us to derive the  anomalous dimension
in the RG evolution equation
\begin{eqnarray}
{d \over d \ln \mu } \Phi_{\rm G}(\omega_1, \omega_2, \mu)
&=& - \int_{-\infty}^{+\infty} d \omega_1^{\prime} \int_{-\infty}^{+\infty}   d \omega_2^{\prime} \,
\Phi_{\rm G}(\omega_1^{\prime}, \omega_2^{\prime}, \mu)
\nonumber \\
&& \times \, \Gamma_{\rm G} (\omega_1, \omega_2, \omega_1^{\prime}, \omega_2^{\prime}, \mu)  \,,
\label{RG equation of phi_G}
\end{eqnarray}
by virtue  of the customary relation
\begin{eqnarray}
\Gamma_{\rm G}
&=& \int_{-\infty}^{+\infty} \,  d \omega_1^{\prime \prime}  \, \int_{-\infty}^{+\infty} d \omega_2^{\prime \prime} \,
Z_{\rm G}(\omega_1, \omega_2, \omega_1^{\prime \prime}, \omega_2^{\prime \prime}, \mu)
\nonumber \\
&&  \hspace{2.6 cm} \times \,
{d Z_{\rm G}^{-1}(\omega_1^{\prime \prime}, \omega_2^{\prime \prime}, \omega_1^{\prime}, \omega_2^{\prime},  \mu) \over d \ln \mu}
\nonumber \\
&& + \,  \delta(\omega_1 - \omega_1^{\prime}) \, \delta(\omega_2 - \omega_2^{\prime}) \,
{d \ln  {\cal F}_B(\mu)  \over d \ln \mu}.
\end{eqnarray}
The renormalization-scale dependence of  ${\cal F}_B(\mu)$
has been determined at four loops \cite{Grozin:2023dlk}.
The renormalization constant $Z_{\rm G}$ can be obtained by evaluating the matrix element
of  ${\cal O}_{\rm G}(\omega_1, \omega_2)$ with the partonic external state
$ \langle 0 | {\cal O}_{\rm G}(\omega_1, \omega_2) | \bar q (\omega_1^{\prime}) g(\omega_2^{\prime},  \eta)  h_v \rangle $.
The  variables $\omega_1^{\prime} \equiv n \cdot k_1$ and $\omega_2^{\prime} \equiv \bar n \cdot k_2$ correspond to
the particular light-cone components of the soft quark and gluon momenta,
while $\eta$ represents the polarization vector of the external gluon.

\begin{figure}[tp]
\begin{center}
\includegraphics[width=1.0 \columnwidth]{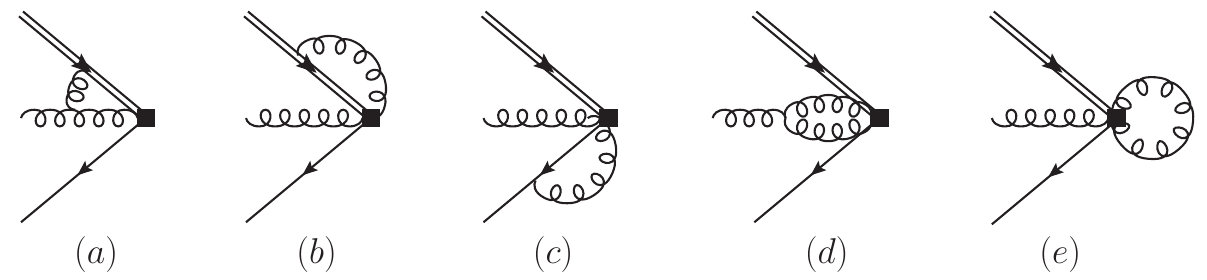}
\caption{Sample Feynman diagrams for evaluating the one-loop QCD correction to the soft function $\Phi_{\rm G}$.
The non-local vertices with two and three gluons include  all possible combinations of the gluon emanating from
the field strength tensor and from the soft Wilson lines $S_n$ and $S_{\bar n}$. }
\label{fig: sample Feynman diagrams for the soft function}
\end{center}
\end{figure}

The sample Feynman diagrams for evaluating  the  renormalization factor $Z_{\rm G}$ at ${\cal O}(\alpha_s)$ are
explicitly displayed in Figure \ref{fig: sample Feynman diagrams for the soft function}.
We adopt dimensional regularization with space-time dimension $d=4-2 \epsilon$ to extract the UV singularities
and implement off-shell regularization to isolate the infrared (IR) divergences of the considered partonic matrix element.
In analogy to the ${\cal O}(\alpha_s)$ correction to $\phi_B^{+}(\omega, \mu)$ \cite{Grozin:1996pq,Lange:2003ff,Braun:2003wx,Bell:2008er},
the soft-gluon exchange between the effective heavy quark and the light quark does not yield the UV-divergent contribution
in Feynman gauge which is employed throughout our calculation.
In addition, the one-gluon exchange between the light quark and the external gluon field is UV finite,
on the basis of the power-counting analysis of the transverse-momentum $\ell_{\perp}$ integration.
Moreover, connecting  one soft gluon from $G_{\mu \nu}(\tau_2 \bar n)$ to the external bottom quark
generates a vanishing correction, since the yielding integrand is an odd function of $\ell_{\perp}$.
The one-particle irreducible diagram from attaching the single gluon field of  $G_{\mu \nu}(\tau_2 \bar n)$
to the external light quark  also brings about the UV finite effect,
in terms of the power-counting analysis of the $\ell_{\perp}$ integral.

The most distinctive feature  of  $\Gamma_{\rm G}$ stems from
the  diagram (e) in Figure \ref{fig: sample Feynman diagrams for the soft function}
comprising of  four independent pieces:
I) both two gluons in the loop from the soft Wilson lines while the external gluon  from the field strength tensor;
II) the two gluons in the loop from the Wilson lines along the $n$ direction and $G_{\mu \nu}(\tau_2 \bar n)$
while the external gluon again from the field strength tensor;
III) the two gluons in the loop from the Wilson lines along the $n$ direction and $G_{\mu \nu}(\tau_2 \bar n)$
while the external gluon from the  Wilson lines;
IV)  both two gluons in the loop from $G_{\mu \nu}(\tau_2 \bar n)$
while  the external gluon  from the soft Wilson lines.
The third type of the ${\cal O}(\alpha_s)$ correction  vanishes,
because the obtained integrand of the $\ell_{\perp}$ integral turns out to be an odd function.
Contracting the two gluon fields from the non-Abelian term in the  field strength tensor
(namely, the fourth type  mechanism) evidently generates a vanishing contribution.
Moreover, the second class of the QCD correction  is cancelled
by the relevant contribution from the diagram (d).
The remaining type-I correction  from the diagram (e) can be cast in the form
\begin{eqnarray}
&&  Z_{\rm G}^{(\rm e)}    \supset    {\alpha_s   \over 4 \pi}  \, {C_A  \over \epsilon}  \,
\bigg \{  \left [ {1 \over \epsilon} +  \ln {\mu^2  \over \omega_1 \omega_2 - i 0}
+  i \, \pi \, \theta(\omega_1 \omega_2)  \right ]
\nonumber \\
&& \hspace{2.5 cm} \times \, \delta(\omega_1 - \omega_1^{\prime}) \, \delta(\omega_2 - \omega_2^{\prime})
 + \left ({i \over 2 \, \pi} \right ) \,
\\
&&    \left [ H_{+}(\omega_1, \omega_1^{\prime}) - H_{-}(\omega_1, \omega_1^{\prime})
- 2    i  \pi\delta(\omega_1 - \omega_1^{\prime}) \, \theta(\omega_2^{\prime} - \omega_2)  \right ]
\nonumber \\
&&
\left [ H_{+}(\omega_2, \omega_2^{\prime}) - H_{-}(\omega_2, \omega_2^{\prime})
- 2    i  \pi\delta(\omega_2 - \omega_2^{\prime}) \, \theta(\omega_1^{\prime} - \omega_1)  \right ] \bigg \},
\nonumber
\end{eqnarray}
where we have introduced the two primitive kernels
$H_{\pm}(\omega_i, \omega_i^{\prime}) =  \theta( \pm \omega_i)  F^{> (<)} (\omega_i, \omega_i^{\prime})
+  \theta( \mp \omega_i)  G^{< (>)} (\omega_i, \omega_i^{\prime})$ as defined in \cite{Beneke:2022msp}
(see also  \cite{Boer:2023vsg} for an overview).
The superscripts ``$>$" and ``$<$" characterize  the positive and negative light-cone momentum $\omega_i$, respectively.
The manifest expressions for  $F^{> (<)}$ and $G^{> (<)}$ read
\begin{eqnarray}
F^{>}(\omega_i, \omega_i^{\prime}) & = &
\left [ \frac{\omega_i \, \theta (\omega_i^{\prime} - \omega_i)} {\omega_i^{\prime} \,  (\omega_i^{\prime} - \omega_i)}   \right ]_{+}
+ \left [ \frac{\theta (\omega_i - \omega_i^{\prime})}  {\omega_i - \omega_i^{\prime}}  \right ]_{\oplus} \,,
\nonumber \\
F^{<}(\omega_i, \omega_i^{\prime})  & = &
\left [ \frac{\omega_i \, \theta (\omega_i - \omega_i^{\prime})} {\omega_i^{\prime} \,  (\omega_i- \omega_i^{\prime})}   \right ]_{+}
+ \left [ \frac{\theta (\omega_i^{\prime} - \omega_i)}  {\omega_i^{\prime} - \omega_i}  \right ]_{\ominus}  \,,
\\
G^{>}(\omega_i, \omega_i^{\prime})  & = & (\omega_i + \omega_i^{\prime}) \,
\left [ \frac{\theta (\omega_i^{\prime} - \omega_i)} {\omega_i^{\prime} \,  (\omega_i^{\prime} - \omega_i)}   \right ]_{+}
- i  \pi  \delta(\omega_i - \omega_i^{\prime}) \,,
\nonumber \\
G^{<}(\omega_i, \omega_i^{\prime})  & = &   (\omega_i + \omega_i^{\prime}) \,
\left [ \frac{ \theta (\omega_i - \omega_i^{\prime})} {\omega_i^{\prime} \,  (\omega_i- \omega_i^{\prime})}   \right ]_{+}
+ i  \pi \delta(\omega_i - \omega_i^{\prime})  \,.
\nonumber
\end{eqnarray}
The standard definition of the ``+" distribution in the variable $\omega_i^{\prime}$ \cite{Lange:2003ff} has been employed.
We further  introduce the modified $``\oplus"$ and $``\ominus"$ distributions to
regulate the integrals of the  non-local terms   $\theta(\omega_i-\omega_i^{\prime})/(\omega_i-\omega_i^{\prime})$
(with  $\omega_i^{(\prime)} > 0$)
and $\theta(\omega_i^{\prime}-\omega_i)/(\omega_i^{\prime}-\omega_i)$ (with  $\omega_i^{(\prime)}  < 0$) \cite{Beneke:2022msp}
\begin{eqnarray}
&& \int_{-\infty}^{+\infty} d \omega_i^{\prime} \, \left [ f(\omega_i,  \omega_i^{\prime}) \right ]_{\oplus/\ominus} \,
\varphi(\omega_i^{\prime})
\nonumber \\
&& = \int_{-\infty}^{+\infty} d \omega_i^{\prime} \, f(\omega_i,  \omega_i^{\prime})
\left [ \varphi(\omega_i^{\prime}) - \theta(\pm \omega_i^{\prime}) \, \varphi(\omega_i)  \right ]  \,.
\end{eqnarray}
The modified $``\oplus"$ distribution in $F^{>}(\omega_i, \omega_i^{\prime})$ generates an interesting
pattern of  evolving  the negative $\omega_i^{\prime} $ into the positive $\omega_i$,
while the modified $``\ominus"$  distribution in $F^{<}(\omega_i, \omega_i^{\prime})$
yields the novel mixing from $\omega_i^{\prime} >0$ to $\omega_i<0$ as already noticed in \cite{Beneke:2022msp}.
Consequently, the support region of $\Phi_{\rm G}(\omega_1, \omega_2, \mu)$ must be extended to the entire real axes
$- \infty < \omega_{1, 2} < + \infty$.

The next-to-leading-order (NLO) contributions from the two diagrams (a) and (b)
with  $\omega_{1, 2}^{\prime} >0$  can be  extracted from
the counterpart expressions for the  twist-three bottom-meson LCDA $\Phi_3(\omega_1, \omega_2, \mu)$  \cite{Offen:2009mt},
by invoking the  exchange  symmetry of  $n \leftrightarrow \bar n$
in the diagrammatic computations (only valid at ${\cal O}(\alpha_s)$ accuracy).
The intriguing UV divergences in the negative support region from these two diagrams
are captured by the modified $``\oplus"$ functions
and by the emerged  $\theta(-\omega_i)$ terms with the standard $``+"$ distributions.
The UV divergent contributions of the diagram (c) arise from
attaching the gluon field of the  Wilson line in the $n$ direction to the external light quark
(while the external gluon state from the field strength).
The yielding UV divergences in the positive support region can be inferred from the corresponding result
of  the leading-twist $B$-meson LCDA \cite{Lange:2003ff,Bell:2008er}.
The one-loop renormalization constant from this diagram in the negative support region
can be obtained by implementing the replacement rules $\omega_1 \to - \omega_1$ and
$\omega_1^{\prime} \to - \omega_1^{\prime}$ in the determined  expression at $\omega_1>0$ and $\omega_1^{\prime}>0$.

Collecting all the individual pieces together, we can readily derive the one-loop anomalous dimension
\begin{widetext}
\begin{eqnarray}
\Gamma_{\rm G} &=&  { \alpha_s \over \pi} \,
\bigg \{ \left [ C_F  \left ( \ln{\mu \over \omega_1 - i 0} - {1 \over 2} \right )
+  C_A  \left ( \ln{\mu \over \omega_2 - i 0} + {i \over 2}  \pi  \right )   \right ]
\delta(\omega_1 - \omega_1^{\prime})  \delta(\omega_2 - \omega_2^{\prime})
-  C_F  H_{+}(\omega_1, \omega_1^{\prime})  \delta(\omega_2 - \omega_2^{\prime})
\nonumber \\
&&  -  \, C_A  H_{+}(\omega_2, \omega_2^{\prime}) \delta(\omega_1 - \omega_1^{\prime})
+ C_A  \left ( {\omega_2 \over \omega_2^{\prime  \, 2}}  \right )
\left [ \theta(\omega_2)  \theta(\omega_2^{\prime} - \omega_2)
-  \theta(- \omega_2)  \theta(\omega_2 - \omega_2^{\prime})  \right ]
\delta(\omega_1 - \omega_1^{\prime}) +  \Delta \Gamma_{\rm G} \bigg \},
\nonumber \\
\Delta \Gamma_{\rm G} &=&  {i \over 4} {C_A \over \pi}
\left [ H_{+}(\omega_1, \omega_1^{\prime}) - H_{-}(\omega_1, \omega_1^{\prime})
- 2 i \pi \delta(\omega_1 -  \omega_1^{\prime}) \right ] \,
 \left [ H_{+}(\omega_2, \omega_2^{\prime}) - H_{-}(\omega_2, \omega_2^{\prime})
- 2 i \pi \delta(\omega_2 -  \omega_2^{\prime})  \right ].
\label{1-loop anomalous dimension}
\end{eqnarray}
\end{widetext}
It is straightforward to verify  that  the  peculiar  terms in  (\ref{1-loop anomalous dimension})
with the colour factor $C_F$ for $\omega_1 >0$ and $\omega_1^{\prime} >0$
(apart from an overall factor of  $\delta(\omega_2 - \omega_2^{\prime})$)
recovers the well-known Lange-Neubert kernel of the twist-two $B$-meson LCDA \cite{Lange:2003ff}.
We further note that the one-loop anomalous dimension (\ref{1-loop anomalous dimension})
becomes complex due to the  soft-parton rescattering,
in analogy to the earlier observation on the QED-generalized bottom-meson soft functions \cite{Beneke:2022msp}.
In contrast with the evolution  kernel for 
$\Phi_3(\omega_1, \omega_2, \mu)$ \cite{Offen:2009mt},
the one-gluon exchange between the light quark and the external gluon field in the UV renormalization of $\Phi_{\rm G}$
cannot generate the structure $\delta(\omega_1+\omega_2-\omega_1^{\prime}-\omega_2^{\prime})$
due to the momentum conservation in the common light-cone direction.
In addition, the emerged structure of   (\ref{1-loop anomalous dimension}) implies   that
the counterpart coordinate-space evolution kernel could  be expressed in terms of
the generator of special conformal transformations $K = v^{\mu} K_{\mu}$  \cite{Braun:2019wyx} potentially.
An elegant construction of this evolution kernel  with the conformal symmetry technique
\cite{Braun:2003rp,Braun:2014owa,Braun:2019wyx}
is an important topic on its own and we plan to investigate this interesting issue in our subsequent work.

\section{The analytic solution}

We are now in a position to derive an exact solution to the RG evolution equation
of the generalized $B$-meson soft function with the  Laplace (or Mellin) transform technique \cite{Lee:2005gza,Bell:2013tfa,Ball:2008fw,Li:2012md,Li:2013xna}.
The resulting  RG equation in Laplace space  becomes local in the first two arguments
and can then be solved analytically.
It turns out to be more advantageous to divide the support of the soft function $\Phi_{\rm G}$
into four separate  pieces
\begin{eqnarray}
\Phi_{\rm G}(\omega_1, \omega_2, \mu)
&=& \theta(\omega_1) \,  \theta(\omega_2) \, \Phi^{>, \, >}_{\rm G}(\omega_1, \omega_2, \mu)
\nonumber \\
&&  + \, \theta(\omega_1) \,  \theta(- \omega_2) \, \Phi^{>, \, <}_{\rm G}(\omega_1, \omega_2, \mu)
\nonumber \\
&&  + \,  \theta(- \omega_1) \,  \theta(\omega_2) \, \Phi^{<, \, >}_{\rm G}(\omega_1, \omega_2, \mu)
\nonumber \\
&& + \, \theta(- \omega_1) \,  \theta(- \omega_2) \, \Phi^{<, \, <}_{\rm G}(\omega_1, \omega_2, \mu)   \,.
\hspace{0.8 cm}
\label{decomposition of the soft function}
\end{eqnarray}
Implementing the Laplace transform for these new soft functions $\Phi^{>(<), \, >(<)}_{\rm G}$
with respect to the two variables $\ln (\mu / \omega_1)$ and $\ln (\mu / \omega_2)$ leads to
\begin{eqnarray}
\tilde{\Phi}^{>(<), \,  >(<)}_{\rm G}(\eta_1, \eta_2, \mu)
& \equiv &  \int_{0}^{\infty} \, {d \omega_1  \over \omega_1} \, \int_{0}^{\infty} {d \omega_2  \over \omega_2}
\left ( { \mu \over \omega_1}  \right )^{\eta_1}
\nonumber \\
&& \hspace{-2.0 cm}
\left ( { \mu \over \omega_2}  \right )^{\eta_2} \,
\Phi^{>(<), \, >(<)}_{\rm G}(\pm \omega_1, \, \pm \omega_2, \, \mu)  \,,
\label{Laplace transform of the soft function}
\end{eqnarray}
where the upper and lower signs in front of $\omega_{1, \, 2}$ correspond to
the superscripts ``$>$" and ``$<$", respectively.
Subsequently,  we obtain a coupled system of the first-order differential equations
for $\tilde{\Phi}^{>(<), \, >(<)}_{\rm G}$   in Laplace space  (with $-1  < {\rm Re} (\eta_{1, 2}) < 0$)
\begin{eqnarray}
&& \left (  {d  \over d \ln \mu} - \eta_1 - \eta_2 \right ) \, {\bf \tilde{\Phi}}_{\rm G}(\eta_1, \eta_2, \mu)
 \\
&& =
- {\alpha_s(\mu) \over \pi} \,
\left [ \tilde{\Gamma}_{\rm G}^{(0)}
+  \tilde{\Gamma}_{\rm G}^{(1)}
+ \left ( {i\over 4 \pi} \right )   \,  \tilde{\Gamma}_{\rm G}^{(2)} \right ] \,
{\bf \tilde{\Phi}}_{\rm G}(\eta_1, \eta_2, \mu) \,,
\nonumber
\label{RGE system in Lapace space}
\end{eqnarray}
where ${\bf \tilde{\Phi}}_{\rm G}$ stands for the column vector
$(\tilde{\Phi}^{>, \, >}_{\rm G} \,,  \tilde{\Phi}^{>, \, <}_{\rm G} \,,
\tilde{\Phi}^{<, \, >}_{\rm G} \,, \tilde{\Phi}^{<, \, <}_{\rm G})^{\rm T}$.
The  evolution matrices  $\tilde{\Gamma}_{\rm G}^{(i)}$ ($i=0,\, 1,  \, 2$) are given by
\begin{eqnarray}
\tilde{\Gamma}_{\rm G}^{(0)} &=&  \left [ C_F \, \left (\partial_{\eta_1} - {1 \over 2} \right )
+ C_A \,  \left (  \partial_{\eta_2}  + {1 \over 1 - \eta_2} +  {i \over 2} \, \pi  \right )  \right ] \,,
\nonumber \\
\tilde{\Gamma}_{\rm G}^{(1)} &=& C_F \, \tilde{\Gamma}_{\rm G}^{(+)}(\eta_1)  \otimes  \mathbf{1}_{\rm 2 \times 2}
+  C_A \, \mathbf{1}_{\rm 2 \times 2} \otimes  \tilde{\Gamma}_{\rm G}^{(+)}(\eta_2)   \,,
 \nonumber  \\
\tilde{\Gamma}_{\rm G}^{(2)} &=&   C_A \, \tilde{\Gamma}_{\rm G}^{(+-)}(\eta_1)
 \otimes  \tilde{\Gamma}_{\rm G}^{(+-)}(\eta_2) \,.
\end{eqnarray}
For brevity, we have introduced the following conventions
\begin{eqnarray}
\tilde{\Gamma}_{\rm G}^{(+)}(\eta) &=&
\left(
  \begin{array}{cc}
   H_{\eta} + H_{-\eta}   \,\,  &  \Gamma(\eta)\, \Gamma(1-\eta) \\
   0  & \,\,  H_{-1-\eta} + H_{-\eta}    \\
  \end{array}
\right),
 \\
\tilde{\Gamma}_{\rm G}^{(+-)}(\eta) &=&
\left(
  \begin{array}{cc}
   H_{\eta} - H_{-1-\eta} + i\, \pi   \,\,  &  \Gamma(\eta)\, \Gamma(1-\eta) \\
   - \Gamma(\eta)\, \Gamma(1-\eta)  & \,\,  H_{-1-\eta} - H_{\eta} + i\, \pi    \\
  \end{array}
\right),
\nonumber
\end{eqnarray}
where  $H_{\eta}$ is the Harmonic number function.
Both the diagonal and non-diagonal terms  of the evolution kernel for $\tilde{\Phi}^{>, \, >}_{\rm G}$ ($\tilde{\Phi}^{<, \, <}_{\rm G}$)
with the colour factor $C_F$  appear to be  identical to  the corresponding expressions
for the QED-generalized distribution amplitude $\tilde{\Phi}_{>}$  ($\tilde{\Phi}_{<}$) \cite{Beneke:2022msp}.

Diagonalizing the one-loop anomalous dimension matrix \cite{Buras:1991jm} leads to the general solution in Laplace space
\begin{eqnarray}
{\bf \tilde{\Phi}}_{\rm G}(\eta_1, \eta_2, \mu)
=  {\rm exp} \left [ V + 2 \, \gamma_E \, (a_1 + a_2) \right ]  \,
\left ( {\mu \over \mu_0} \right )^{\eta_1 + \eta_2}   \,
&& \nonumber \\
 \hat{U}^{-1}(\eta_1,  \eta_2) \,\,  {\rm diag} \left (1, \, 1, \, 1, \, e^{- i \pi a_2}   \right )  \,
\, \hat{U}(\eta_1+a_1,  \eta_2+a_2)
&&  \nonumber  \\
\times \,  \frac{\Gamma(1-\eta_1) \, \Gamma(1+ \eta_1 + a_1) }{\Gamma(1 + \eta_1) \, \Gamma(1 -  \eta_1 - a_1)}   \,
\frac{\Gamma(2-\eta_2) \, \Gamma(1+ \eta_2 + a_2) }{\Gamma(1 + \eta_2) \, \Gamma(2 -  \eta_2 - a_2)}   \,
&& \nonumber \\
\times \, {\bf \tilde{\Phi}}_{\rm G}(\eta_1+a_1, \eta_2+a_2, \mu_0),
\hspace{3.8 cm} &&
\label{Solution to the RGE in Lapace space}
\end{eqnarray}
where we have introduced the evolution functions  \cite{Bell:2013tfa,Lee:2005gza}
\begin{eqnarray}
V &=&  - \int_{\mu_0}^{\mu} {d \mu^{\prime} \over \mu^{\prime}}
\frac{\alpha_s( \mu^{\prime}) }{\pi}
\nonumber \\
&& \times \, \bigg [ C_F  \left ( \ln{\mu^{\prime} \over \mu_0} - {1 \over 2} \right )
 + C_A   \left ( \ln{\mu^{\prime} \over \mu_0} +  {i  \over 2}  \pi \right ) \bigg ],
 \\
a_1 &=&  - \int_{\mu_0}^{\mu} {d \mu^{\prime} \over \mu^{\prime}} \,
\frac{\alpha_s( \mu^{\prime})}{\pi}  \, C_F  \,,
\hspace{0.2  cm}
a_2 =  - \int_{\mu_0}^{\mu} {d \mu^{\prime} \over \mu^{\prime}} \,
\frac{\alpha_s( \mu^{\prime})}{\pi}  \, C_A  \,.
\nonumber
\end{eqnarray}
The  transformation matrix $\hat{U}$ in the obtained  solution
(\ref{Solution to the RGE in Lapace space}) can be written as
\begin{eqnarray}
\hat{U}(\eta_1,  \eta_2)  = \left(
  \begin{array}{cc}
   1   \,\,  &  -e^{-i \pi \eta_1}  \\
   1   \,\,  &  -e^{i \pi \eta_1}    \\
  \end{array}
\right)
\otimes
\left(
  \begin{array}{cc}
   1   \,\,  &  -e^{-i \pi \eta_2}  \\
   1   \,\,  &  -e^{i \pi \eta_2}    \\
  \end{array}
\right)
\,.  \hspace{0.5 cm}
\end{eqnarray}
Carrying out the inverse Laplace  transformation with respect to the two variables $\eta_1$ and  $\eta_2$,
we then  derive the desired momentum-space solution to the RG equation (\ref{RG equation of phi_G})
\begin{eqnarray}
{\bf{\Phi}}_{\rm G}(\omega_1, \omega_2, \mu)
&=& {\rm exp} \left [ V + 2  \gamma_E  (a_1 + a_2) \right ]
\int_{0}^{\infty} { d \omega_1^{\prime} \over \omega_1^{\prime} }
\int_{0}^{\infty} { d \omega_2^{\prime} \over \omega_2^{\prime} }
\nonumber \\
&& \hspace{-1.5 cm}
\left ( {\mu_0 \over \omega_1^{\prime}} \right)^{a_1}
\left ( {\mu_0 \over \omega_2^{\prime}} \right)^{a_2}
\left  [ {\cal J}_{1}   \otimes  {\cal J}_{2}
+ \frac{1 - e^{-i  \pi  a_2}}{4} \,  {\cal K}_{1}   \otimes  {\cal K}_{2} \right ]
\nonumber \\
&& \hspace{-1.5 cm}
\times \, {\bf  \Phi}_{\rm G}(\omega_1^{\prime}, \omega_2^{\prime}, \mu_0),
\label{Final RG solution in momentum space}
\end{eqnarray}
where 
${\bf  \Phi}_{\rm G}$ represents the  column vector with the elements
defined  by  the four momentum-space soft functions with the particular arguments
on the right-hand side of (\ref{Laplace transform of the soft function}).
The evolution matrices ${\cal J}_{i}$ and ${\cal K}_{i}$ (with $i=1, \, 2$)
can be expressed in terms of the familiar  Meijer-${\rm G}$ functions
and we do not present their lengthy expressions here.

\section{Phenomenological implications}

In order to explore the asymptotic behaviours of the bottom-meson  soft function
at  small and large quark and gluon momenta,
we will take the exponential model with only the positive domain for the initial condition
\begin{eqnarray}
\Phi_{\rm G}^{\rm exp}(\omega_1, \omega_2, \mu_0) &=&  \frac{\lambda_E^2 + \lambda_H^2}{6} \,
{\omega_1 \, \omega_2^2  \over \omega_0^5} \,
{\rm exp} \left (- {\omega_1 + \omega_2 \over \omega_0} \right ) \,
\nonumber \\
&& \times \, \theta(\omega_1) \, \theta(\omega_2) \,,
\label{exponential model of phiG}
\end{eqnarray}
at $\mu_0 = 1.0 \, {\rm GeV}$, motivated by the non-perturbative analysis of the HQET sum rules
 at leading order in $\alpha_s$ \cite{Qin:2022rlk}.
The hadronic quantities $\lambda_E^2$ and $\lambda_H^2$ are defined by the effective matrix elements
of the chromoelectric and  chromomagnetic operators \cite{Grozin:1996pq,Braun:2017liq}.
It remains important to remark that  the model-independent constraints from the operator-product-expansion (OPE) computation
of the regularized  moments \cite{Lee:2005gza,Feldmann:2014ika,Beneke:2023nmj}  can be further implemented
for constructing the  QCD improved model of this soft function.
In accordance with the general solution (\ref{Final RG solution in momentum space})
and the asymptotic expansion of the   Meijer-${\rm G}$ functions,
we can determine the scaling behaviour of the $B$-meson soft function in the endpoint region  $\omega_{1, 2} \to 0$
\begin{eqnarray}
&& \Phi_{\rm G}^{\rm exp}(\omega_1 \to 0 , \omega_2 \to 0, \mu)
\nonumber \\
&& =   \frac{\lambda_E^2 + \lambda_H^2}{6} \,
{1  \over \omega_0^2} \, {\rm exp} \left [ V + 2 \, \gamma_E \, (a_1 + a_2) \right ]
\left ( {\mu_0 \over \omega_0} \right)^{a_1 + a_2}
\nonumber \\
&&  \hspace{0.5 cm}
\frac{1 - e^{-i  \pi  a_2}}{4 \pi^2}   \Gamma(1+a_1)\Gamma(1+a_2)
+  {\cal O} \left ({\omega_1 \over\omega_0 },  {\omega_2 \over\omega_0 } \right ).
\hspace{0.5 cm}
\label{asymptotic behaviour of phiG at small momenta}
\end{eqnarray}
Interestingly, this soft function acquires a constant value in the limit $\omega_{1, 2} \to 0$,
bearing a resemblance to the counterpart asymptotic behaviour of
the QED-generalized LCDA $\Phi_{0-}(\omega, \mu)$ \cite{Beneke:2022msp}.

\begin{figure}[tp]
\begin{center}
\includegraphics[width=0.80 \columnwidth]{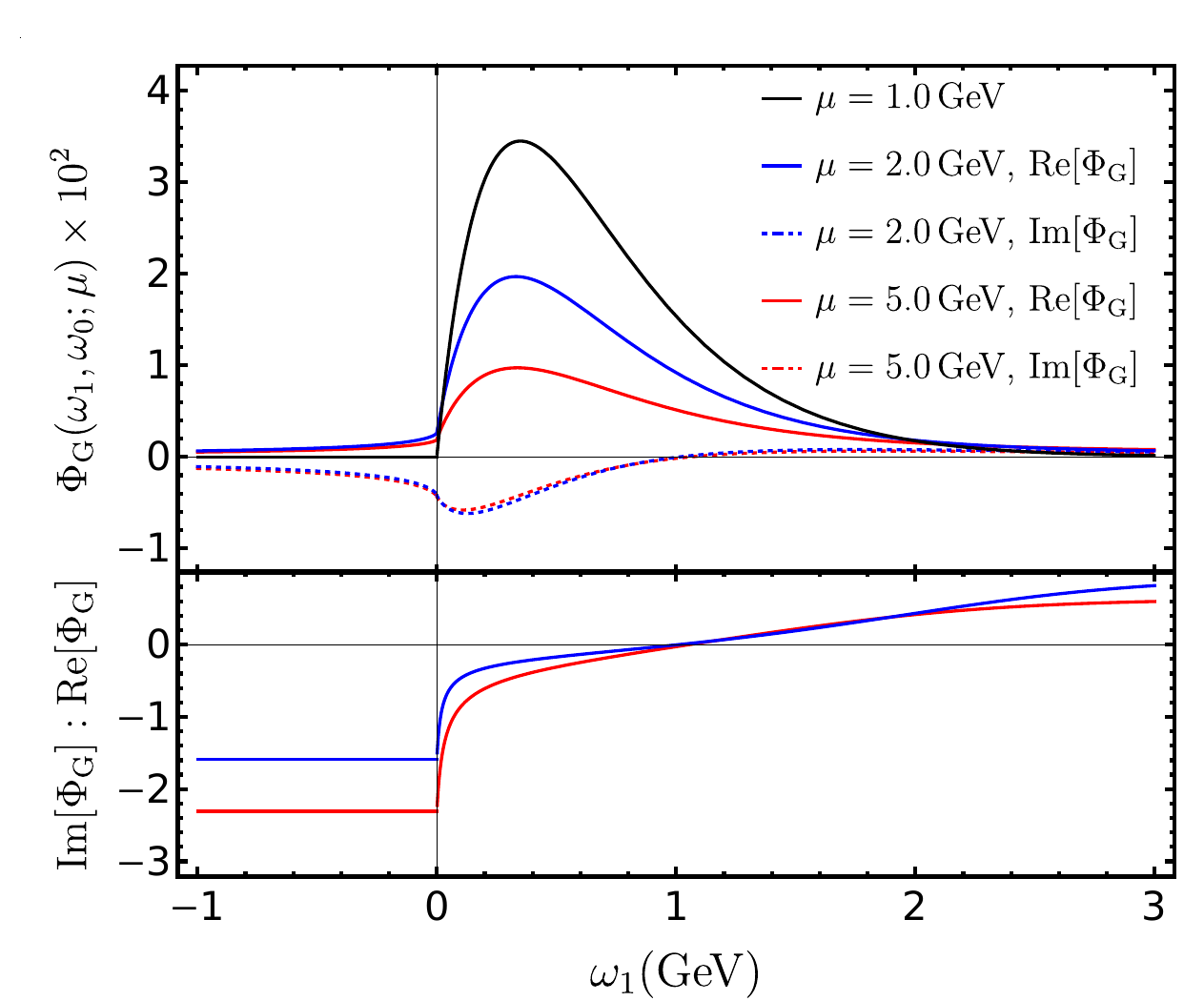}
\vspace*{0.1cm}
\includegraphics[width=0.80  \columnwidth]{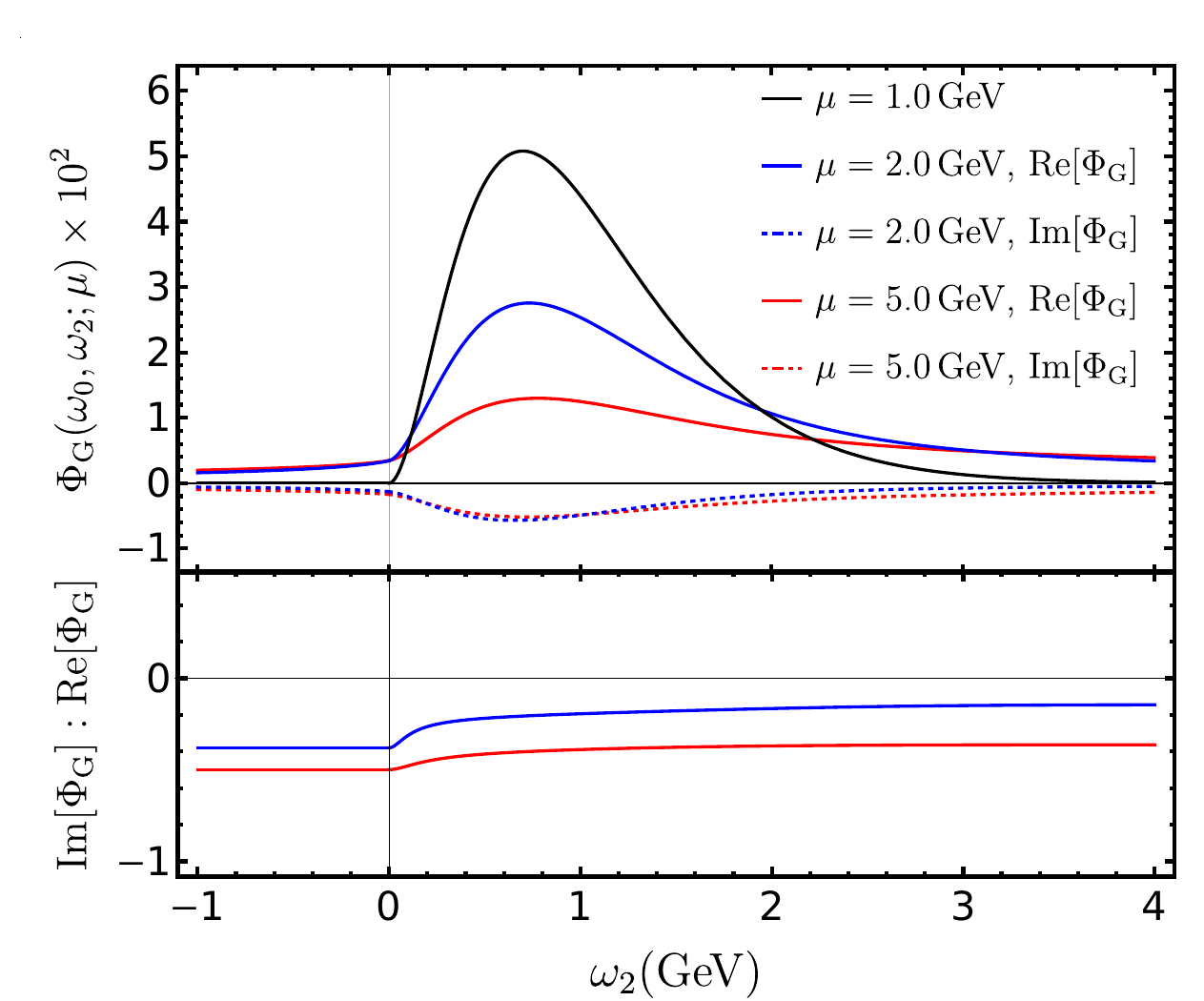}
\caption{Theory predictions for the RG evolution of $\Phi_{\rm G}(\omega_1, \omega_0, \mu)$ (upper panel)
and $\Phi_{\rm G}(\omega_0, \omega_2, \mu)$  (lower panel) by taking the  exponential model (\ref{exponential model of phiG})
as an initial condition (black curves).
The blue and red curves are obtained by evolving the soft functions to a hard-collinear scale $\mu=2.0 \, {\rm GeV}$
and to a hard  scale $\mu=5.0 \, {\rm GeV}$, respectively.
The ratios between the imaginary and real parts of the evolved soft functions are also shown explicitly.
We have employed the numerical values of $\omega_0$, $\lambda_E^2$ and $\lambda_H^2$
from \cite{Shen:2020hfq}.}
\label{fig: numerical effect of the RG evolution}
\end{center}
\end{figure}

Along the same vein, we can further  derive the large-momentum  behaviour of $\Phi_{\rm G}$ with the solution
(\ref{Final RG solution in momentum space})
\begin{eqnarray}
&& \Phi_{\rm G}^{\rm exp}(\omega_1 \to \pm \infty, \omega_2 \to \pm \infty, \mu)
\nonumber \\
&& \propto   \frac{\lambda_E^2 + \lambda_H^2}{6} \,
{1  \over \omega_0^2} \,
{\rm exp} \left [ V + 2 \, \gamma_E \, (a_1 + a_2) \right ]  \,
\left ( {\mu_0 \over \omega_0} \right)^{a_1 + a_2}  \,
\nonumber \\
&& \hspace{0.5 cm}
\left ( {\omega_0 \over \pm \omega_1} \right )^{1+a_1} \,
\left ( {\omega_0 \over \pm \omega_2} \right )^{1+a_2}  \,,
\label{asymptotic behaviour of phiG at large momenta}
\end{eqnarray}
which turns out to be analogous to the well-known behaviour of the leading-twist
$B$-meson LCDA \cite{Lange:2003ff}.
We are then led to conclude that all non-negative moments of the soft function
$\Phi_{\rm G}(\omega_1, \omega_2, \mu)$ are divergent.
The non-existence of the normalization integral can also be understood from
the logarithmic UV singularities of the HQET matrix element on the left-hand side
of (\ref{definition of phi_G}) at $\tau_{1, 2} \to 0$
(see \cite{Braun:2003wx} for the earlier discussion).
Adopting the three-parameter  ans\"{a}tz for  the  bottom-meson soft function
 in \cite{Qin:2022rlk}, we  can further verify analytically that the determined scaling  behaviours at $\omega_{1, 2} \to 0$
and  $\omega_{1, 2} \to \pm \infty$ remain unchanged for this more general model.

To develop a transparent understanding of the RG evolution effect due to the one-loop anomalous dimension,
we display in Figure \ref{fig: numerical effect of the RG evolution}
the numerical predictions for $\Phi_{\rm G}$
evolved to the two distinct renormalization scales $\mu=2.0 \, {\rm GeV}$ and $5.0 \, {\rm GeV}$.
It is evident that incorporating the one-loop QCD evolution can
bring about the considerable reduction of the soft function in the peak region
at $\mu=2.0 \, {\rm GeV}$ (as large as ${\cal O} (50 \, \%)$ numerically).
Moreover, the RG evolution can result in the positive correction in the large $\omega_{1, \, 2}$ region
due to the emergence of the radiative tail which falls off much slower than the initial behaviour (\ref{exponential model of phiG}).
In particular,  taking into account  the leading-logarithmic (LL) evolution of the soft function $\Phi_{\rm G}$
can generate the noticeable imaginary part as anticipated: approximately  ${\cal O} (20 \, \%)$ of the corresponding real part,
even if we start with  the real-valued initial condition  (\ref{exponential model of phiG}).
This new strong-phase source will be  of  importance
for predicting the CP-violating observables  in the  exclusive bottom-meson decays.
The RG evolution effect from  $\mu_0 = 1.0 \, {\rm GeV}$
to   $\mu = 5.0 \, {\rm GeV}$ becomes more pronounced than
the observed pattern at $\mu=2.0 \, {\rm GeV}$.
The numerical features of the evolution effects in Figure \ref{fig: numerical effect of the RG evolution}
further indicate the constant phases of $\Phi_{\rm G}(\omega_1, \omega_0, \mu)$
and $\Phi_{\rm G}(\omega_0, \omega_2, \mu)$ in the negative domain,
which  remain true for an arbitrary real-valued initial condition
factorizable in the two momentum variables with positive support.

\section{Conclusions}

In conclusion,  we have derived for the first time the RG evolution equation
of the generalized $B$-meson distribution amplitude,
defined by the subleading  HQET matrix element with non-aligned partonic fields.
The most striking feature of the UV renormalization  kernel of the soft function
consists in  the novel pattern of mixing positive into negative support, irrespective of the initial behaviour of  $\Phi_{\rm G}$,
thus extending the support region of this generalized  distribution amplitude  to the entire real axes
$- \infty < \omega_{1, 2} < + \infty$.
Adopting one sample  model for the $B$-meson soft function,
we have explicitly demonstrated that including  the RG evolution from the hadronic scale  $\mu_0 = 1.0 \, {\rm GeV}$
to the hard-collinear scale $\mu=2.0 \, {\rm GeV}$ can  result in an enormous reduction of $\Phi_{\rm G}$ in the peak region.
Extending our  RG analysis  to the generic soft functions defined with partonic fields localized
on two distinct light-cone directions will be highly beneficial for  evaluating the long-distance charming penguin contributions
in the $B \to K^{(\ast)} \ell^{+} \ell^{-}$ and $B \to K^{\ast} \gamma$ decays.

%
\begin{acknowledgments}
\section*{Acknowledgements}

Y.J. acknowledges support from the Deutsche Forschungsgemeinschaft (DFG, German Research Foundation)
through the Sino-German Collaborative Research Center TRR110 ``Symmetries and the Emergence of Structure in QCD''
(DFG Project-ID 196253076, NSFC Grant No. 12070131001, - TRR 110).
The research of Y.L.S. is supported by the  National Natural Science Foundation of China  with
Grant No. 12175218 and the Natural Science Foundation of Shandong with Grant No.  ZR2020MA093.
C.W. is supported in part by the National Natural Science Foundation of China
with Grant No. 12105112 and  the Natural Science Foundation of
Jiangsu Education Committee with Grant No. 21KJB140027.
Y.M.W. acknowledges support from the  National Natural Science Foundation of China  with
Grant No.  12075125.

\end{acknowledgments}

\bibliographystyle{apsrev4-1}

\bibliography{References}

\end{document}